\documentclass[11pt]{article}
\usepackage{amssymb}
\usepackage{epsfig}
\newlength{\bredde}
\def\slash#1{\settowidth{\bredde}{$#1$}\ifmmode\,\raisebox{.15ex}{/}
\hspace*{-\bredde} #1\else$\,\raisebox{.15ex}{/}\hspace*{-\bredde} #1$\fi}
\newcommand{\be}{\begin{equation}}
\newcommand{\ee}{\end{equation}}
\newcommand{\bea}{\begin{eqnarray}}
\newcommand{\eea}{\end{eqnarray}}
\newcommand{\nn}{\nonumber}

\newcommand{\al}{\alpha}
\newcommand{\la}{\lambda}
\newcommand{\ga}{\gamma}
\newcommand{\sect}[1]{\setcounter{equation}{0}\section{#1}}

\def\wa{w^{(a)}}
\def\fat{f^{(a)}(\tau)}
\def\zc{{z^\ast}}

\def\tr{{\mbox{tr}}}
\def\re{{\Re\mbox{e}}}
\def\im{{\Im\mbox{m}}}

\begin{document}
\topmargin -1.4cm
\oddsidemargin -0.8cm
\evensidemargin -0.8cm
\title{\Large{{\bf 
The solution of a chiral random matrix model with complex eigenvalues
}}}

\vspace{1.5cm}

\author{~\\{\sc G. Akemann}\\~\\
Service de Physique Th\'eorique\footnote{Unit\'e associ\'ee 
CNRS/SPM/URA 2306}, CEA/Saclay\\
F-91191 Gif-sur-Yvette Cedex, France
}
\date{}
\maketitle
\vfill
\begin{abstract}
We describe in detail the solution of the extension of the chiral Gaussian 
Unitary Ensemble (chGUE) into the complex plane. The correlation functions of 
the model are first calculated for a finite number of $N$ complex eigenvalues, 
where we exploit the existence of orthogonal Laguerre polynomials in the 
complex plane. When taking the large-$N$ limit we derive new correlation 
functions in the case of weak and strong non-Hermiticity, thus describing 
the transition from the chGUE to a generalized Ginibre ensemble. Applications 
to the Dirac operator eigenvalue spectrum in QCD with non-vanishing chemical 
potential are briefly discussed. This is an extended version of hep-th/0204068.
\end{abstract}
\vfill

\begin{flushleft}
SPhT T02/053\\
\end{flushleft}
\thispagestyle{empty}
\newpage

\renewcommand{\thefootnote}{\arabic{footnote}}
\setcounter{footnote}{0}

\sect{Introduction}\label{intro}

The application of Random Matrix Models (RMM) in a variety of physical 
systems has a long history \cite{GMW}. Roughly speaking its applications can 
be divided into two classes. RMM can be either used to describe local 
properties. The correlations for example of energy levels 
often shown a universal behavior on the distance of the mean level spacing,
making a RMM description possible.
Choosing the RMM with the same global symmetries as the underlying 
Hamiltonian of the system 
fixes completely the form of the local spectral fluctuations. 
On the other hand RMM can be used as a combinatorial tool to count classes 
of graphs on random surfaces, which may then be interpreted as Feynman graphs 
of a corresponding Quantum Field Theory. 
Our investigations here belong to the first class of applications on a local
level, 
and the corresponding correlation functions are often called microscopic. 

In most cases RMM have to be seen as effective models owing their validity 
from symmetry considerations alone. However, in some cases the mapping to the 
theory one wishes to describe can be made exact, as for example in 
dynamical triangulations of random surfaces and its relation to Quantum 
Gravity \cite{Ambbook}
or in the application of RMM to the Dirac operator spectrum in four 
dimensional QCD \cite{DOTV}. 
At the present stage the RMM we propose here as a model for
QCD with chemical potential is just an effective 
or phenomenological model. But a more precise mapping to QCD 
via chiral perturbation theory may also be possible.

In many applications models of random 
matrices having real eigenvalues are considered. 
All such possible models have been classified \cite{Zirn} 
and all the corresponding correlation functions are known (for a complete list
of results and references see e.g. \cite{Ivanov}). On the other hand there are 
also physical situations where complex eigenvalues occur. 
As examples we mention localization in superconductors \cite{HN}, 
dissipation and scattering in Quantum Chaos \cite{GHS} 
or Quantum Chromodynamics (QCD) with a non-vanishing 
chemical potential \cite{Steph}.
Other possible applications include the fractional Quantum Hall Effect 
as in \cite{PdF} or a two-dimensional plasma of charged particles 
\cite{Janco,FJ}.
A classification of all models with non-Hermitian matrices 
has been achieved more recently \cite{Denis}
and the number of possible cases exceeds by far those with real eigenvalues. 
But even when considering the simplest extension of the three 
RMM classes introduced by Dyson \cite{Dyson} much less is known. They consist 
of real symmetric ($\beta=1$), complex Hermitian ($\beta=2$) and quaternion 
self-dual matrices ($\beta=4$) where the Dyson index $\beta$ labels the 
ensemble by the power of the Jacobian diagonalizing the 
matrices,  which is given by the Vandermonde determinant $\Delta$.
Due to their invariance these ensembles are called Gaussian Orthogonal,
Unitary or Symplectic Ensemble, the GOE, GUE and GSE respectively.

In a seminal work the extension of these RMM into the complex plane has been 
obtained by Ginibre \cite{Gin}. He calculated the joint probability 
distribution function (j.p.d.f.) for $\beta=2$ and $4$ and partially for 
$\beta=1$ as well as all microscopic correlation functions for $\beta=2$. 
Since then progress has been slow. The full j.p.d.f. for $\beta=1$ was 
calculated much more recently in \cite{Edelman} 
but the correlation functions are still lacking 
up to date. The correlation functions for $\beta=4$ were found in 
\cite{Mehta}. 
All these models are Gaussian ensembles with independently distributed matrix 
elements. More recently generalized Gaussian ensembles were considered 
containing a parameter $\tau$ that permits to interpolate between real 
and fully complex eigenvalues. 
The global density interpolating between the Wigner semi-circle on the real 
line and the constant density on a circular disc of Ginibre was found in 
\cite{So} (see also \cite{HOV} for numerical results 
of all three ensembles).
Together with the introduction of the interpolating parameter $\tau$ the 
regime of so-called weak non-Hermiticity was discovered \cite{FKS,Ef} 
for $\beta=2$ and $1$. For $\beta=2$ it 
interpolates smoothly between the know microscopic correlations of the 
GUE and those of Ginibre called strong non-Hermitian limit \cite{FKS}. 
The techniques to derive such results are either supersymmetry 
\cite{FKS,Ef} or orthogonal polynomials \cite{FKS98,FJ} 
which also exist on the complex plane \cite{PdF}.  
For $\beta=4$ the $\tau$-dependent correlations were discovered very recently 
in \cite{EK} in both the weak and strong limit. 
New correlation functions for $\beta=2$ were also found in \cite{A01} where
so-called Dirac mass terms were introduced in the application to 
three-dimensional QCD with complex eigenvalues. 

From the RMM \cite{SV} 
which was first proposed to describe the QCD Dirac spectrum of 
real eigenvalues, the chiral Gaussian Unitary Ensemble (chGUE), 
we know that the model we are aiming for should belong to the class of chiral
models. In the presence of a quark chemical potential it is also known that 
the Dirac eigenvalues become complex \cite{Steph}. 
We thus wish to define and solve 
such a chiral RMM with complex eigenvalues. The model is found by taking 
the know analogy between the GUE and the chGUE on the level of real 
eigenvalues 
and translating it into the complex plane. The starting point is the known 
complex GUE introduced in \cite{PdF}. In analogy to the real case and from 
the finding of Hermite polynomials in the complex plane for the non-chiral 
complex GUE in \cite{PdF} we expect and do find Laguerre polynomials in the 
complex plane. This enables us to provide explicit solutions at a finite 
number $N$ of eigenvalues as well as in the weak and strong non-Hermitian 
large-$N$ limit.

The situation we wish to describe, QCD in the presence of a chemical 
potential, is quite delicate. The difficulties encountered when trying
to do first principle calculations by using Monte Carlo simulations on 
a lattice are reviewed in 
\cite{lattmu}. The are due to the problem of convergence 
for a determinant of a Dirac operator with complex eigenvalues. 
Despite the fact that recent progress was made in \cite{Fodor} 
the full phase diagram with $\mu\neq 0$ remains unexplored so far.
The analytical knowledge 
of microscopic correlation functions from 
complex chiral RMM could thus be very useful, in view of its success 
in predicting real eigenvalue correlations.

The chemical potential problem has already been addressed in quite 
a few works using RMM. In \cite{Steph} the failure of the 
quenched approximation was illuminated by Stephanov. 
He also calculated the global eigenvalue 
density and its envelope as a functions of $\mu$. We will use his results 
to determine our non-Hermiticity parameter $\tau$ 
as a function of $\mu$ for small 
values of $\mu$. In \cite{Halasz} the temperature-density phase diagram 
of QCD with two light flavors was predicted using a RMM. A more refined model 
incorporating also vector and axial symmetries of the action was studied in  
\cite{Benoit}. Effective models for a phase transition with the spectral 
density vanishing to an arbitrary power
were introduced in \cite{ADMNII,AVII}. Furthermore, 
the analyticity properties of the partition function \cite{HOSV}, problems of 
convergence \cite{HJV} as well as the global symmetry of such Dirac operators 
\cite{HOV} were analyzed using RMM. 
Finally first data for quenched simulations of QCD with $\mu\neq 0$ have been 
confronted with RMM predictions, finding a crossover from the Wigner-Dyson 
class over 
strongly non-Hermitian, Ginibre type correlations to a Poisson distribution  
\cite{MPW}. 
There the nearest neighbor spacing of eigenvalues in the bulk 
from non-chiral RMM were compared to the data.
Here, we wish to provide analytic 
information on all microscopic correlation functions 
at the origin, which are important for chiral symmetry breaking \cite{BC}, 
using a chiral eigenvalue model. 

The paper is organized as follows. In section \ref{model} we 
introduce a chiral complex eigenvalue model. A justification is given why to 
call it chiral and why in particular it represents a complex extension 
of the chGUE. The identification of our parameter $\tau$ 
as a function of the chemical potential $\mu$  
in QCD is made by comparing with the RMM \cite{Steph}.
In the next section \ref{OPchap} we introduce the orthogonal polynomials 
belonging to our model, show that they are given by the Laguerre polynomials 
in the complex plane and give the exact solution for all correlators for a 
finite number of eigenvalues $N$. Sections \ref{weak} and \ref{strong} treat 
the large-$N$ limit at weak and strong non-Hermiticity, respectively. 
New correlation functions are found in both cases by taking appropriate 
limits of the Laguerre polynomials. We illustrate our findings with 
a few examples in each case. We also prove that the weak correlations 
interpolate between the known correlations of the chGUE and those in the 
strong limit from section \ref{strong}. 
This is done by letting the weak non-Hermiticity parameter 
$\al$ approach zero or infinity in sections \ref{weak} and \ref{strong} 
respectively, explicitly checking that the correlations coincide. 
Our conclusions are presented in the last section \ref{con}.
A brief summary of our results has been already published in \cite{A02}.


\sect{The model}\label{model}

We first define our model in terms of complex eigenvalues $z_{j=1,\ldots,N}$ 
and then explain in detail why it is the generalization of the chGUE into 
the complex plane. The partition function is defined as 
\be
{\cal Z}_{chiral}^{(a)}(\tau) \ \equiv\  \int \prod_{j=1}^N
\left(dz_j dz_j^\ast\ \wa(z_j)\right)
\left|\Delta_N(z_1^2,\ldots,z_N^2)\right|^2, 
\label{Zev}
\ee
where we have introduced the following weight function
\be
\wa(z)=|z|^{2a+1}\! 
\exp\left[
-\frac{N}{1-\tau^2}\left(|z|^2 -\frac{\tau}{2}(z^2+\zc^2)\right)
\right], 
\label{weight}
\ee 
and $\Delta_N(z_1,\ldots,z_N)=\prod_{k>l}^N (z_k-z_l)$ is 
the Vandermonde determinant. 
The parameter $a>-1$ can be taken real  for the time being. In the application 
to QCD it will be an integer, $a=N_f+\nu$, where $N_f$ labels the number of 
massless quark flavors and $\nu$ is the sector of topological charge.
The parameter $\tau\in[0,1]$ controls the degree of non-Hermiticity. 
For $\tau=1$ we are back to the chGUE with real eigenvalues, as the weight 
functions (\ref{weight}) then reduces to a $\delta$-function in the imaginary 
part $\im(z)$ times a Gaussian weight for the real part $\re(z)$. 
On the other hand for $\tau=0$ 
the degree of non-Hermiticity is maximal, and we will see the individual 
r\^{o}le of $\tau$ in the weight of the real and imaginary part below.

The motivation for the definition of the model (\ref{Zev}) comes from a 
comparison to the non-chiral model of complex eigenvalues, the complex 
extension of the GUE, introduced in \cite{PdF,FKS,A01}. 
Let us therefore recall the relation between the GUE and chGUE for real 
eigenvalues first.
The chGUE is defined in terms of a complex matrix $W$ of size $N\times(N+\nu)$ 
\cite{SV}
\bea
{\cal Z}_{chGU\!E}^{(N_f,\nu)} &=& \int dWdW^\dagger \det(WW^\dagger)^{N_f} 
\mbox{e}^{-Ng\,\tr(WW^\dagger)} = 
\int_0^\infty \prod_{j=1}^N \left( d\la_j\la_j^{N_f+\nu} 
\mbox{e}^{-Ng\la_j}\right)|\Delta_N(\la_1,\ldots,\la_N)|^2 
\nn\\ 
&=& \int_{-\infty}^\infty \prod_{j=1}^N 
\left( dx_j |x_j|^{2(N_f+\nu)+1} \mbox{e}^{-Ng\,x_j^2} \right) 
|\Delta_N(x_1^2,\ldots,x_N^2)|^2 \ .
\label{ZchGUE}
\eea
In the first step we have diagonalized and given the partition function in 
terms of the positive eigenvalues\footnote{To be precise these are called 
singular values.} $\la_j$ of the matrix $W$ while in the second step we have 
simply substituted $\la_j=x_j^2$. Note the squares inside the Vandermonde 
due to this transformation. Now we compare to the GUE given in terms of a 
Hermitian matrix $H$ of size $N\times N$, 
\be
{\cal Z}_{GU\!E}^{(2N_f)} = \int dH \det(H)^{2N_f} 
\mbox{e}^{-Ng\,\tr(H^2)} 
= \int_{-\infty}^\infty \prod_{j=1}^N 
\left( dx_j x_j^{2N_f} \mbox{e}^{-Ng\,x_j^2} \right) 
|\Delta_N(x_1,\ldots,x_N)|^2 \ .
\label{ZGUE}
\ee
Both models can be solved using the method of orthogonal polynomials 
\cite{Mehta} and the corresponding sets are given by the generalized 
Laguerre polynomials 
for the chGUE and by the Hermite polynomials for the GUE (at $N_f=0$, 
for $N_f>0$ see \cite{DN}). Moreover, when taking the parameter $a=N_f+\nu$ 
to be real we can reconstruct the orthogonal polynomials of 
the chGUE on the full real line from the even subset of the 
orthogonal polynomials of the GUE. Because of the squares 
in the Vandermonde in eq. (\ref{ZchGUE}) we only need even polynomials
there. This relation also holds for more general, non-Gaussian weight 
functions (see e.g. in \cite{ADMNII}). 
For $a=-1/2$ this is nothing else than the known relation between Hermite and 
Laguerre polynomials \cite{grad} (see eq. (\ref{HeLrel}) below). 

Let us now turn to the complex generalization of the GUE as introduced in 
\cite{PdF,FKS,A01}. The Hermitian matrix $H$ is replaced by a complex matrix
\be
J\ =\ H + i \sqrt{\frac{1-\tau}{1+\tau}} A \ .
\label{Jdef}
\ee
with $\tau\in[0,1]$. 
The weight for both the Hermitian and anti-Hermitian part of $J$, the 
Hermitian matrices $H$ and $A$
respectively, is chosen to be Gaussian with variance $(1+\tau)/2N$. 
We therefore obtain \cite{FKS}
\bea
{\cal Z}^{(2N_f)}(\tau)&\equiv& \int dJdJ^{\dagger}
\det(JJ^\dagger)^{N_f}
 \exp\left[-\frac{N}{1-\tau^2}\,\tr\left(JJ^\dagger-\frac{\tau}{2}(
     J^2+(J^\dagger)^2)\right)\right]
\label{ZPdF}\\
&=&
 \int \prod_{j=1}^N 
\left(dz_jdz_j^\ast\  |z_j|^{2N_f} \exp\left[
-\frac{N}{1-\tau^2}\left(|z_j|^2 -\frac{\tau}{2}(z_j^2+z_j^{\ast2})\right)
\right]\right)
 |\Delta_N(z_1,\ldots,z_N)|^2 \ .
\nn
\eea
For averages over the matrix elements of $J$ this corresponds to
\be
\langle J_{kl}J_{lk}\rangle \ =\ \frac{\tau}{N} \ , \ \ \ 
\langle J_{kl}J_{kl}^\ast\rangle \ =\ \frac{1}{N} \ ,
\ee
which illustrates the interpolating r\^{o}le of $\tau$:
for $\tau=1$ we have $J_{lk}=J_{kl}^\ast$ and thus Hermitian matrices
whereas for $\tau=0$ all $J_{kl}$ are mutually independent
corresponding to maximal non-Hermiticity.
In the latter case the model was already defined and solved by Ginibre 
in \cite{Gin} where the orthogonal polynomials are monomials in the complex
eigenvalues $z_j$ of $J$. Here, we can study the transition 
between the GUE and the 
Ginibre ensemble and the corresponding polynomials were found to be  
Hermite polynomials in the complex plane \cite{PdF}. 
Using the same reasoning as in the mapping between the real 
eigenvalues model of the 
GUE eq. (\ref{ZGUE}) and the chGUE eq. (\ref{ZchGUE}), eq. 
(\ref{ZPdF}) immediately leads to our model eq. (\ref{Zev}) above.
This way we have solved the problem of how to extend the positive real 
eigenvalues (\ref{ZchGUE}) into the complex plane. 
We may of course ask ourselves to what kind of matrix representation 
eq. (\ref{Zev}) would correspond to. 
We would have to express eq. (\ref{Zev}) in terms 
of a different matrix $\tilde{J}$, probably with 
two different complex matrices now, in the form 
$\tilde{J}\sim WW^\dagger + i VV^\dagger$. 
However, we have not been able to find such a representation so far.

Let us compare to a different matrix model which has been 
introduced by Stephanov 
for QCD with non-vanishing chemical potential \cite{Steph},
\be
{\cal Z}^{(N_f,\nu)} (\mu)\ \equiv\  \int dWdW^\dagger 
\det\left(
\begin{array}{cc}
0&iW+\mu\\
iW^\dagger+\mu&0\\
\end{array}
\right)^{N_f} 
\mbox{e}^{-Ng\,\tr(WW^\dagger)} \ .
\label{Zsteph}
\ee
Here, the matrix inside the determinant replaces the QCD Dirac operator 
with the chemical potential added as $\ga_0\mu$. 
The disadvantage from adding a constant 
matrix to the chiral random matrix is that the model can no longer be written 
in terms of eigenvalues of the matrix $W$. 
Our idea is to use the model eq. (\ref{Zev}) instead and 
to suppose that on the scale of the mean level spacing of the complex
Dirac operator eigenvalues 
the microscopic correlations are the same. 
A somewhat similar observation has been made in \cite{GW} where a chiral 
matrix model with temperature was considered. In order to capture the effect 
of the lowest Matsubara frequency a constant matrix $i\ga_0 t$ was added to 
the Dirac matrix. 
An analysis in terms of eigenvalues of the matrix $WW^\dagger$ 
is also no longer possible, although the original 
Dirac eigenvalues remain real in this case. 
However, the microscopic correlations remain the same compared to the model 
at $t=0$, when translating to the same mean level spacing \cite{GW}. 
To avoid misunderstandings let us stress that we do not claim that the 
correlations at $\mu=0$ and $\mu\neq 0$ are the same but that  
(\ref{Zsteph}) at $\mu\neq 0$ and our model at $\tau<1$ share the same 
microscopic correlation functions. 

In order to relate these two parameters we compare the macroscopic spectral 
densities of the two models. 
In our model it can be read off from \cite{So} since the 
preexponential factors in the weight eq. (\ref{weight}) 
are subdominant in the macroscopic large-$N$ limit,
\be
\rho_\tau(z)= \frac{1}{\pi(1-\tau^2)}\ , \ \ 
\ \mbox{if} \ \ \ \ \frac{x^2}{(1+\tau)^2}+\frac{y^2}{(1-\tau)^2}\leq 1 \ ,
\label{rhomacro} 
\ee
where $z=x+iy$.
The density is constant and bounded by an ellipse, where for $\tau=0$ we 
recover the result of \cite{Gin}. 
On the other hand for eq. (\ref{Zsteph}) the macroscopic 
density has been calculated 
in \cite{Steph},
\be
\rho_\mu(z)= \frac{1}{4\pi}\left(\frac{\mu^2+y^2}{(\mu^2-y^2)^2} -1\right), 
\label{rhomacrosteph}
\ee
where the eigenvalues are bounded by the following curve\footnote{In our 
conventions the Dirac eigenvalues are real for $\mu=0$, as it 
should have become clear by now. In comparison to ref. \cite{Steph} this
means that we interchange real and imaginary part, $x\leftrightarrow y$, 
compared to there.} 
\be
0= y^4(x^2+4\mu^2)\ +\ y^2(1+2\mu^2(2-x^2-4\mu^2))\ +\ \mu^4(x^2-4(1-\mu^2))\ .
\label{bound}
\ee
The aim of the two models (\ref{Zev}) and (\ref{Zsteph}) is to describe 
the fluctuations of Dirac eigenvalues close to the origin, because of 
their relevance for chiral symmetry breaking from the Banks-Casher relation 
\cite{BC}. We will therefore restrict ourselves to small values of 
$\mu$ since otherwise the determinant in eq. (\ref{Zsteph}) is completely 
dominated by $\mu$. Another reason for small $\mu$ is that we wish to stay 
in the phase with broken chiral symmetry. For $\mu\geq1$ the curve 
(\ref{bound}) splits into two and the vanishing of the macroscopic spectral 
density $\rho_\mu(0)$ 
indicates through the Banks-Casher relation that the chiral phase 
transition has taken place. If we solve the biquadratic equation 
(\ref{bound}) for the imaginary part $y$ and expand in $\mu$ we obtain 
\be
y^2\ =\ \mu^4(4-x^2) \ +\ {\cal O}(\mu^6) \ \ .
\ee
Expanding the density eq. (\ref{rhomacrosteph}) in $\mu\ll 1$ 
we can therefore neglect $y^2$
and finally arrive at
\be
\rho_\mu(z)= \frac{1}{4\pi\mu^2}\ , \ \ 
\ \mbox{if} \ \ \ \ \frac{x^2}{4}+\frac{y^2}{4\mu^2}\leq 1 \ .
\label{rhomu}  
\ee
Comparing eqs. (\ref{rhomacro}) and (\ref{rhomu}) 
we identify
\be
4\mu^2 \ \equiv\ (1-\tau^2) \ ,
\label{mutau}
\ee
which is valid for small chemical potential $\mu$ and the 
parameter $\tau$ close to unity meaning small
non-Hermiticity. We have now identified all the parameters in our model with 
the corresponding quantities in QCD. In contrast to the model (\ref{Zsteph})
by Stephanov we will not be able to study the chiral phase transition since 
our $\rho_\tau(z)$ will remain constant for all $\tau\in[0,1)$. 
On the other hand we will be able to calculate all correlation functions, 
for finite-$N$ as well in two different microscopic scaling limits in the 
next sections. This has not been achieved so far for the model  (\ref{Zsteph}),
where only the macroscopic density is known \cite{Steph}.

We finally comment on the presence of the absolute value in the 
weight eq. (\ref{weight}) which is chosen to make the partition function 
real. In the model (\ref{Zsteph}) no absolute value of the Dirac determinant 
is present. Let us however emphasize that also in the 
real eigenvalue model eq. (\ref{ZchGUE})
an absolute value is present due to the mapping from the positive to the 
full real line, even for zero quark flavors $N_f=0$.


\sect{The Solution from Orthogonal Polynomials}\label{OPchap}

We now turn to the solution of our model for a finite number of eigenvalues 
$N$, where we use the powerful method of orthogonal polynomials \cite{Mehta}. 
In the two following sections we then take the large-$N$ limit. 
The orthogonal polynomials in the complex plane we are searching for 
are defined with respect to the weight function eq. (\ref{weight}), 
\be
\int dz dz^\ast\  \wa (z)\ 
P_k^{(a)}(z)P_l^{(a)}(\zc) \ =\ \delta_{kl} \ .
\label{OP}
\ee
All $k$-point correlation functions defined as 
\be
\rho_N^{(a)}(z_1,\ldots,z_k) \ \equiv\ 
\int \prod_{j=k+1}^N\!dz_j dz_j^\ast\ 
\prod_{l=1}^N\wa(z_l)\ 
\left|\Delta_N(z_1^2,\ldots,z_N^2)\right|^2, 
\label{rhodef}
\ee
can be expressed in terms of the kernel of orthogonal polynomials 
\be
K^{(a)}_N(z_1,z_2^\ast)\ \equiv\  
[\wa(z_1)\wa(z_2^\ast)]^{\frac12}
\sum_{k=0}^{N-1} P_k^{(a)}(z_1)P_k^{(a)}(z_2^\ast)  \ .
\label{kernel}
\ee
Due to the invariance of the determinant the monomial powers inside the 
Vandermonde $\Delta_N$ in eq. (\ref{Zev})
can be replaced by the orthogonal polynomials and we arrive at \cite{Mehta}
\be
\rho_N^{(a)}(z_1,\ldots,z_k) \ =\
\det_{1\leq i,j\leq k}\left[K_N^{(a)}(z_i,z_j^\ast)\right].
\label{MM}
\ee
In eq. (\ref{Zev}) only even powers of $z_j$ occur inside the 
Vandermonde $\Delta_N$ and we thus only need even polynomials. 
From the known relation between Hermite and Laguerre polynomials \cite{grad} 
for $a=-1/2$ we expect to obtain Laguerre polynomials of the squared 
argument, $P_k^{(a)}(z)\sim  L_k^a(z^2)$, and we will find that this is 
indeed the case (see eq. (\ref{Laguerre}) below).

In order to derive the orthogonality of the Laguerre polynomials in the 
complex plane we define the normalization integral
\bea
\fat &\equiv&\int dzdz^\ast \  \wa (z) \label{fdef}\\
&=& N^{-a-\frac32}\pi\ \Gamma\left(a+\frac32\right)
(1-\tau^2)^{\frac{a}{2}+\frac34} 
P_{a+\frac32}\left(\frac{1}{\sqrt{1-\tau^2}}\right) \nn
\eea
where $P_{a+\frac32}(x)$ is the Legendre function. While the expression 
given is exact for finite and infinite $N$ it will simplify in the weak 
non-Hermiticity limit in section \ref{weak} where $1-\tau^2$ is rescaled with 
$N$.
Following the idea of ref. \cite{PdF} we use the invariance of the 
integral eq. (\ref{fdef}) under reparametrization to make the 
generating function of the Laguerre polynomials appear. 
Performing a change of variables $z\to \mbox{e}^{i\varphi}z$ inside 
the integral eq. (\ref{fdef}) we obtain from eq. (\ref{weight})
\bea
\fat &=&\int dzdz^\ast  |z|^{2a+1} \exp\left[
-\frac{N}{1-\tau^2}\left(|z|^2 -\frac{\tau}{2}(\mbox{e}^{i2\varphi}z^2
+\mbox{e}^{-i2\varphi}z^{\ast2})\right)\right]\nn\\
\Longleftrightarrow \ \ \ \ 1&=&
\left\langle 
\exp\left[ \frac{(\mbox{e}^{i2\varphi}-1)\tau^2}{1-\tau^2}
\left(\frac{Nz^2}{2\tau}\right)\right]
\exp\left[ \frac{(\mbox{e}^{-i2\varphi}-1)\tau^2}{1-\tau^2}
\left(\frac{N\zc^2}{2\tau}\right)\right]
\right\rangle ,
\label{preLrel}
\eea
where the average is taken with respect to the weight function eq. 
(\ref{weight}) and we have divided by $\fat$ in order to normalize.
We now wish to obtain the following generating functional for the Laguerre 
polynomials, which is also valid for complex arguments $Z$:
\be
(1-u)^{-a-1} \exp\left[ \frac{u}{u-1}\,Z^2\right]
\ =\ \sum_{k=0}^\infty  L_k^a(Z^2)\ u^k \ \ ,\ \ \ \ |u|<1\ ,
\label{genL}
\ee
and similarly for complex conjugate arguments. The split into variables 
$u$ and $Z^2$ is not unique in eq. (\ref{preLrel}). We will use the known 
results \cite{PdF} for the Hermite polynomials $H\!e_n$ of the model eq. 
(\ref{ZPdF}) to read off the argument in the case $a=-1/2$ using \cite{abramo} 
\be
H\!e_{\,2k}\left( \sqrt{\frac{N}{\tau}}z\right) \ =\ 
(-1)^k 2^k k!\ L^{-\frac12}_k \left( \frac{Nz^2}{2\tau}\right).
\label{HeLrel}
\ee
We therefore choose $Z^2=\frac{Nz^2}{2\tau}$ and
\be
\frac{u}{u-1} \ \equiv\ \frac{(\mbox{e}^{i2\varphi}-1)\tau^2}{1-\tau^2}\ ,
\label{udef}
\ee
which can be written as 
\be
u\ =\ \frac{\tau^2(1-\mbox{e}^{i2\varphi})}{1-\tau^2\mbox{e}^{i2\varphi}} \ .
\label{u}
\ee
It is complex and satisfies $|u|<1$. We now multiply eq. (\ref{preLrel}) 
with $[(1-u)(1-u^\ast)]^{-a-1}$ to obtain
\be
[(1-u)(1-u^\ast)]^{-a-1} \ = \ \sum_{k,l=0}^\infty 
\left\langle L^a_k \left( \frac{Nz^2}{2\tau}\right)
 L^a_k \left( \frac{Nz^{\ast2}}{2\tau}\right) \right\rangle
u^k u^{\ast\, l} \ ,
\label{genLL}
\ee
where we have inserted the generating functional eq. (\ref{genL}). 
Using the fact that 
\be
uu^\ast \ =\ \frac{4\tau^4\sin^2\varphi}{(1-\tau^2)^2+4\tau^2\sin^2\varphi}
\label{uu*}
\ee
together with 
the definition (\ref{udef}) it is easy to see that the left hand side 
of eq. (\ref{genLL}) only depends on the combination $uu^\ast$ and we 
obtain 
\be
[(1-u)(1-u^\ast)]^{-a-1}\ =\ [1-\tau^{-2}uu^\ast]^{-a-1} 
\ =\ \sum_{k=0}^\infty \frac{\Gamma(a+1+k)}{\Gamma(a+1)k!}\ 
\frac{(uu^\ast)^k}{\tau^{2k}} \ , 
\label{lhs}
\ee
which converges due to $\tau^{-2}uu^\ast<1$ from eq. (\ref{uu*}).
Consequently also the right hand side of eq. (\ref{genLL}) depends only on the 
combination  
$uu^\ast$ and thus the Laguerre polynomials have to be orthogonal. By 
comparing coefficients of eqs. (\ref{lhs}) and (\ref{genLL}) we arrive at 
\be
\left\langle L^a_k \left( \frac{Nz^2}{2\tau}\right)
 L^a_k \left( \frac{Nz^{\ast2}}{2\tau}\right) \right\rangle
\ =\ \delta_{kl} \ \frac{\Gamma(a+1+k)}{\Gamma(a+1)\,k!} \frac{1}{\tau^{2k}}\ .
\label{ortho}
\ee
The orthogonal polynomials from eq. (\ref{OP}) are thus finally given as 
\be
P_k^{(a)}(z)\equiv \left(
\frac{\Gamma(a+1+k)}{\Gamma(a+1)\,k!}
\fat \right)^{-\frac12}\!
(-\tau)^k L_k^a\left(\frac{Nz^2}{2\tau}\right) .
\label{Laguerre}
\ee
The phase factor $(-1)^k$ which is chosen 
arbitrarily is taken such that the relation 
(\ref{HeLrel}) is reproduced at $a=-1/2$. 
All $k$-point correlation functions are therefore completely determined 
inserting the polynomials $P_k^{(a)}(z)$
into eqs. (\ref{kernel}) and (\ref{MM}). We only give the 
expression for the kernel at finite-$N$, 
\be
K^{(a)}_N(z_1,z_2^\ast)\ =\  
[\wa(z_1)\wa(z_2^\ast)]^{\frac12}
\frac{\Gamma(a+1)}{\fat}
\sum_{k=0}^{N-1} \frac{k!}{\Gamma(a+1+k)}\, \tau^{2k}\ 
 L^a_k \left( \frac{Nz_1^2}{2\tau}\right)
 L^a_k \left( \frac{Nz_2^{\ast2}}{2\tau}\right) \ .
\label{kernelN}
\ee
Let us point out that although we have checked our results 
by comparing to the orthogonal Hermite polynomial of the model 
(\ref{ZPdF}) at $a=-1/2$ 
we cannot do the same check with the correlation functions. 
The reason is that in the model (\ref{ZPdF}) also the odd polynomials 
contribute to the correlation functions.
 
In the following sections we will investigate the large-$N$ limit. 
For that purpose usually a different form of the kernel eq. (\ref{kernel}) 
is given. It only contains the polynomials of order $N$ and $N-1$ and is 
derived from the Christoffel-Darboux identity. However, due to the $\tau$ 
dependence of the arguments in eq. (\ref{Laguerre}) this identity no longer 
holds in the complex plane\footnote{I wish to thank E. Kanzieper for 
pointing out this fact to me.}. The modified Christoffel-Darboux identity 
now contains polynomials of all orders,  $0,\ldots,N$. 
A similar fact occurs in the two-matrix model (see e.g. \cite{Bert}).
The large-$N$ limit is no longer governed by the asymptotic polynomials 
alone and we will have to use special properties of the Laguerre polynomials 
to achieve our goal in the next sections.

Before we continue let us explain why we did not include Dirac quark masses 
into our model, in contrast to \cite{A01}. In the non-chiral model
mass terms of the form $\prod_f^{N_f}|z-im_f|^2$ were included in the 
weight function. The corresponding correlation functions were calculated
by relating them to massless correlations, using the fact that the masses can 
be produced from a larger Vandermonde \cite{AK,A01}. 
Using the same technique in 
our model eq. (\ref{Zev}) we would only be able to treat additional terms 
of the form $\prod_f^{N_f}(z^2+m_f^2)(z^{\ast 2}+m_f^2)$ in the weight 
(\ref{weight}). While these terms might be interesting for other applications 
they are not of the right form for QCD mass terms of a complex Dirac matrix. 
For this reason we refrain from deriving such correlation functions here.


\sect{The weak non-Hermiticity limit}\label{weak}

The large-$N$ limit at weak non-Hermiticity was introduced in \cite{FKS} and 
is defined as follows. We take the limit $\tau\to1$ that leads to Hermitian 
matrices at the same time as $N\to\infty$ such that the combination 
\be
\lim_{N\to\infty}N(1-\tau^2)\ \equiv \ \al^2  
\label{aldef} 
\ee
is kept finite. In this limit the macroscopic density eq. (\ref{rhomacro}) 
depends on real eigenvalues only. It becomes the Wigner semi-circle, 
$\rho(x)=\frac{1}{2\pi}\sqrt{4-x^2}$, for the Gaussian weight 
we have chosen, with support $[-2,2]$ 
on the real line only. However, the properly rescaled 
microscopic correlations remain complex 
and are different from those of real eigenvalues as well as from the 
strong non-Hermiticity limit where $\tau<1$. 
Because of the identification we have made between $\tau$ and $\mu$, 
eq. (\ref{mutau}), eq. (\ref{aldef}) invokes to send also $\mu\to 0$ as 
$N\to\infty$ while keeping 
\be
4N\mu^2 \ =\ \al^2\ 
\label{mual}
\ee
fixed\footnote{In ref. \cite{A01} an attempt was made to relate $\mu$ to 
$\al$ by comparing the macroscopic density (\ref{rhomacrosteph}) 
to the average of microscopic eigenvalues and a wrong power was
obtained.}. 
In other words the weak non-Hermiticity parameter $\al^2$ directly measures
the chemical potential in the microscopic scaling limit. 
The rescaling of $\mu$ is similar to that of the quark masses $m_f$, 
however with a different power in $N$. It makes sure that the 
eigenvalues in the Dirac determinant are not completely dominated by 
the masses or the chemical potential (see in eq. (\ref{Zsteph})).  
Namely in the weak 
non-Hermitian limit at the origin we also rescale the complex eigenvalues 
keeping 
\be
N(\re\ z+i\im\ z)\ =\ Nz \ \equiv \xi \ \ \ ,
\label{microweak}
\ee
The matrix size $N$ corresponds to the volume on the lattice.
It has been mentioned in \cite{HJV} that in the RMM eq. (\ref{Zsteph})
the numerical effort to obtain convergence grows exponentially with 
$N\mu^2$. Keeping it fixed here according to eq. (\ref{mual})
should make a comparison to data feasible. 

Eqs. (\ref{aldef}) - (\ref{microweak}) 
define our microscopic origin scaling limit in the 
complex plane at weak non-Hermiticity. 
The kernel eq. (\ref{kernel}) and correlation functions  eq. (\ref{MM})
also have to be rescaled with the mean level spacing
$1/N$ of the eigenvalues. In the definition of the correlation functions 
for each variable $z_j$ we integrate over its real and imaginary part and 
we obtain one power $1/N$ from each integration.
The microscopic kernel and correlation functions are therefore defined as 
follows in the weak non-Hermiticity limit:
\bea
K^{(a)}_S(\xi_1,\xi_2^\ast) &\equiv&  \lim_{N\to\infty}\frac{1}{N^2}\ 
K^{(a)}_N\left(\frac{\xi_1}{N},\frac{\xi_2^\ast}{N}\right) \nn\\
\rho_S^{(a)}(\xi_1,\ldots,\xi_k)&\equiv&  \lim_{N\to\infty}\frac{1}{N^{2k}}\ 
\rho_N^{(a)}\left(\frac{\xi_1}{N},\ldots,\frac{\xi_k}{N}\right).
\label{microrho+ker}
\eea
When looking at the kernel at finite-$N$ eq. (\ref{kernelN}) and the 
appropriate rescaling of the eigenvalues, eq. (\ref{microweak}), we observe 
that we are missing a power in $N$ inside the argument of the Laguerre 
polynomials. The asymptotic of the Laguerre polynomials, 
\be
\lim_{k\to\infty} k^{-a} L_k^a(Z^2) \ =\ 
(kZ^2)^{-\frac{a}{2}} J_a(2\sqrt{kZ^2})
\label{asymp}
\ee
suggests to introduce a variable $t\equiv\frac{k}{N}$ and thus to replace 
the sum in eq. (\ref{kernelN}) by an integral, 
$\sum_{k=0}^{N-1}\to N\int_0^1dt$. In doing so we write everything in terms 
of the variables $t$ and $\xi_j$. Under the integral we can thus replace 
everywhere the Laguerre polynomials by $J$-Bessel functions using eg. 
(\ref{asymp}) since a finite $t=\frac{k}{N}\in(0,1]$ means that $k$ has to 
become large and the point $t=0$ is of measure zero. 

In the following we separately give the asymptotic of each factor in eq. 
(\ref{kernelN}) before stating the final answer. The weight functions contain 
already the right powers of $N$ in the exponential and using eq. (\ref{aldef}) 
we obtain
\be
\lim_{N\to\infty} \wa(z)^{\frac12} \ =\ N^{-a-\frac{1}{2}} |\xi|^{a+\frac12}
\exp\left[-\frac{1}{\al^2}(\im \xi)^2\right]\ .
\label{weakweight}
\ee
To calculate the weak limit of the normalization integral $\fat$ in eq. 
(\ref{fdef}) we need to know the asymptotic of the Legendre functions, given 
by \cite{bate}
\be
\lim_{\tau\to1} P_{a+\frac12}\left(\frac{1}{\sqrt{1-\tau^2}}\right)
\ =
\ \frac{2^{a+\frac12}\Gamma(a+1)}{\sqrt{\pi}\ 
\Gamma(a+\frac32)}\ (1-\tau^2)^{-\frac{a}{2}-\frac14}\ .
\label{Legendre}
\ee
This leads to 
\be
\lim_{\tau\to1} \fat \ =\ \sqrt{2\pi\al^2} N^{-a-2}2^a \Gamma(a+1)\ .
\label{fatweak}
\ee
Under the integral we first replace 
\be
\lim_{k\to\infty} \frac{k!}{\Gamma(a+1+k)} = N^{-a} t^{-a} \ .
\label{prefactor}
\ee
Rewriting eq. (\ref{aldef}) as $\tau^2=1-\al^2/N$ the power in $\tau$ turns 
into an exponential,
\be
\lim_{N\to\infty}\left(\tau^{2k} \ =\ \mbox{e}^{k\ln(\tau^2)}\right)
\ =\ \mbox{e}^{-t\al^2} \ .
\label{tauasymp}
\ee
Finally The Laguerre polynomials are replaced as 
\be
\lim_{k,N\to\infty} L_k^a\left(\frac{Nz^2}{2\tau}\right) \ =\ 
N^a (2t)^{\frac{a}{2}} \xi^{-a} J_a(\sqrt{2t}\xi) \ .
\label{Lasymp}
\ee
Putting together eqs. (\ref{weakweight}) - (\ref{Lasymp}) all powers of $N$ 
cancel as they should and we obtain 
\be
K_S^{(a)}(\xi_1,\xi_2^\ast) =
\frac{|\xi_1\xi_2^\ast|^{\frac12}}{\sqrt{2\pi\al^2}}
\frac{|\xi_1\xi_2^\ast|^a}{(\xi_1\xi_2^\ast)^a}
\mbox{e}^{-\frac{1}{\al^2}
\left((\Im m\xi_1)^2+(\Im m\xi_2^\ast)^2\right)}
\int_0^1 dt\ \mbox{e}^{-\al^2t} 
J_a(\sqrt{2t}\xi_1)J_a(\sqrt{2t}\xi_2^\ast) .
\label{weakkernel}
\ee
When inserting the kernel into the determinant eq. (\ref{MM}) for the 
correlators all factors apart from the integral can be taken out. In 
particular the second factor cancels due to the occurrence of the $\xi_j^{-a}$ 
and $\xi^{\ast-a}_j$ in each column and row. We arrive at the 
microscopic, weakly non-Hermitian correlation functions:
\be
\rho^{(a)}_S(\xi_1,\ldots,\xi_k) \ =\ \prod_{l=1}^k
\left(
\frac{|\xi_l|}{\sqrt{2\pi\al^2}}\ 
\mbox{e}^{-\frac{2}{\al^2}(\Im m\xi_l)^2}
\right)
\det_{1\leq i,j\leq k}\left[
\int_0^1 dt\ \mbox{e}^{-\al^2t} 
J_a(\sqrt{2t}\xi_i)J_a(\sqrt{2t}\xi_j^\ast)
\right]  .
\label{weakrho}
\ee
They are clearly different from the correlations of the non-chiral model eq. 
(\ref{ZPdF}) as calculated for $a=0$ in \cite{FKS} 
and for $a=N_f$ massless flavors in \cite{A01}. There, the product of two 
$J$-Bessel functions is replaced by a single cosine, 
$\cos(\sqrt{2t}(\xi_i-\xi_j^\ast))$ which corresponds to a half-integer 
Bessel $J_{-\frac12}$.
\begin{figure}[-h]
\centerline{\epsfig{figure=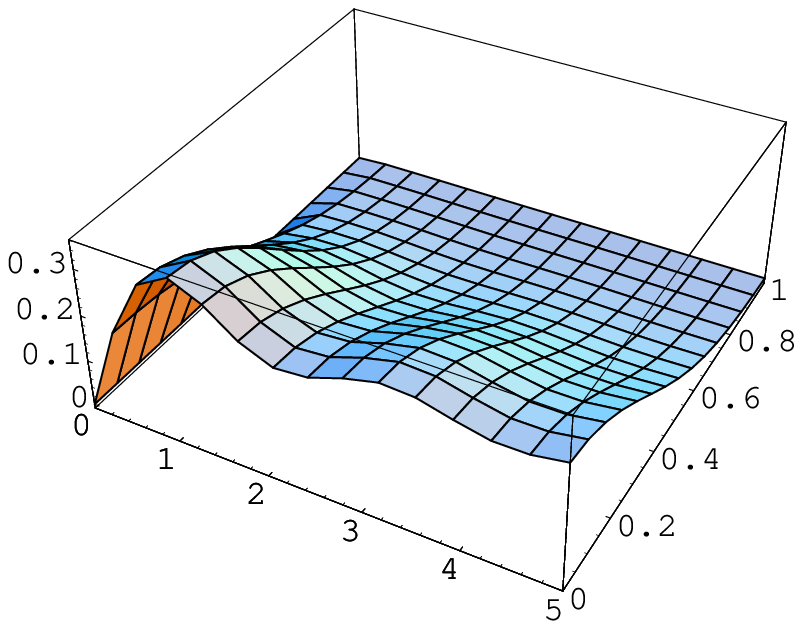,width=20pc}
\put(-160,20){$\re\ \xi$}
\put(-30,40){$\im\ \xi$}
\put(-240,140){$\rho_s^{(0)}(\xi)$}
\put(80,20){$\re\ \xi$}
\put(210,40){$\im\ \xi$}
\put(0,140){$\rho_s^{(1)}(\xi)$}
\epsfig{figure=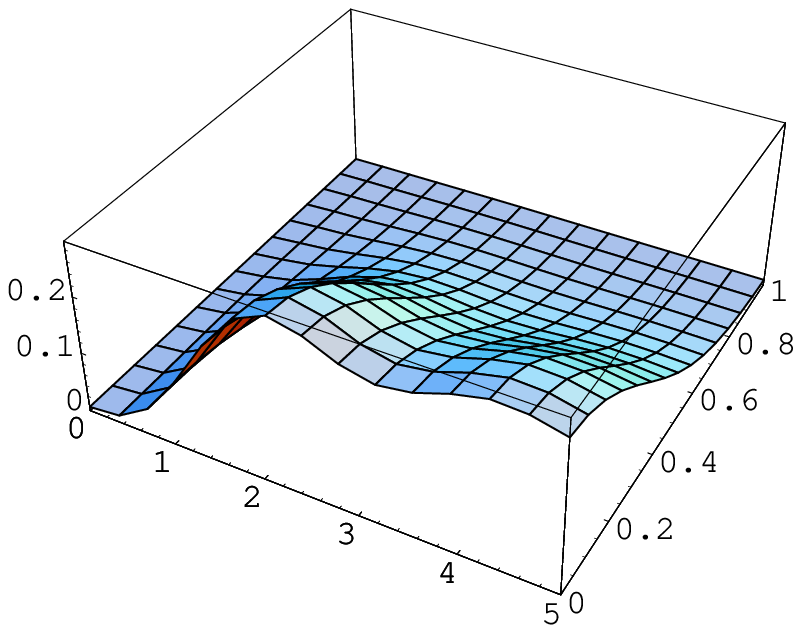,width=20pc}}
\caption{
The microscopic density at $\al^2=0.6$ for $a=0$ (left) and $a=1$ (right) 
massless flavors. 
}
\label{weakplot}
\end{figure}
As an important check we can compare to the correlation functions of real 
eigenvalues \cite{SV} of the chGUE eq. (\ref{ZchGUE}) which have been shown 
to be universal \cite{ADMN}. We take the Hermitian limit by sending 
$\al^2\to 0$ which corresponds to $\tau=1$.
The exponential prefactors become $\delta$-functions for the imaginary parts 
of the eigenvalues, as they should. The integral inside the determinant then 
makes the universal Bessel kernel (of squared arguments)
appear,
reproducing the known correlation functions of \cite{SV}:
\be
\int_0^1dt J_a(\sqrt{2t}x_1)J_a(\sqrt{2t}x_2) \ = \ 
\frac{x_1J_{a+1}(\sqrt{2}x_1)J_a(\sqrt{2}x_2)
-x_2J_{a+1}(\sqrt{2}x_2)J_a(\sqrt{2}x_1)}{\sqrt{2}(x_1^2-x_2^2)}\ .
\ee
To give an example for eq. (\ref{weakrho}) we show 
the microscopic density in the weak limit for two values of $a$ in Fig 
\ref{weakplot}.
For one quark flavor ($a=1$) the level repulsion at the origin is stronger 
as expected.
The plot continues symmetrically from the given first quadrant into the full
complex plane by reflecting along the real and imaginary axis. 
In cutting along the positive real axis we can see the oscillations 
familiar from the correlations of real eigenvalues. Each maximum corresponds 
to the location of a single eigenvalue in the complex plane. 
We also show the $\al$-dependence of the quenched microscopic density in 
Fig. \ref{weakalphadep}. For 
increasing $\al$ and thus for increasing rescaled 
chemical potential $\mu$ the eigenvalues spread 
further into the complex plane. In particular the density also builds up 
along the imaginary axis. As we will see in the next section in the limit 
$\al\to\infty$ we obtain the microscopic correlations at strong 
non-Hermiticity to be defined below. The limiting picture to be compared with 
is Fig. \ref{strongplot} below with the quenched density on the left.
\begin{figure}[-h]
\centerline{\epsfig{figure=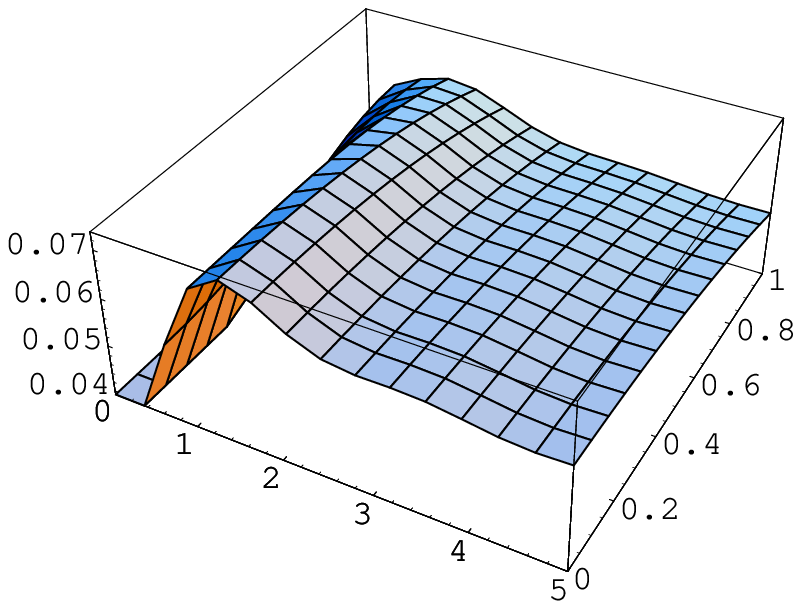,width=20pc}
\put(-160,20){$\re\ \xi$}
\put(-30,50){$\im\ \xi$}
\put(-240,140){$\rho_s^{(0)}(\xi)$}
\put(80,20){$\re\ \xi$}
\put(210,50){$\im\ \xi$}
\put(0,140){$\rho_s^{(0)}(\xi)$}
\epsfig{figure=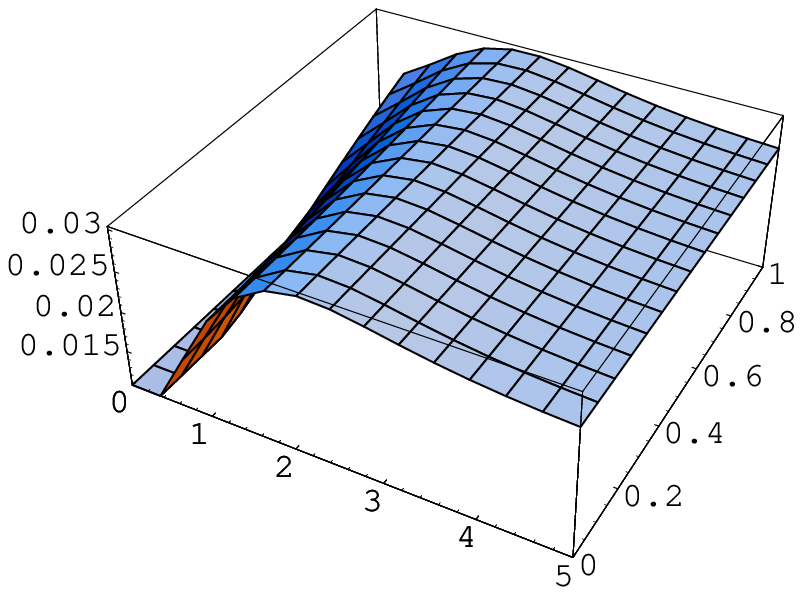,width=20pc}}
\caption{
The quenched microscopic density ($a=0$) at $\al^2=1.6$ 
(left) 
and $\al=2.5$ (right).
}
\label{weakalphadep}
\end{figure}


\sect{The strong non-Hermiticity limit}\label{strong}

We now turn to the strong non-Hermiticity limit. In this limit $\tau\in[0,1)$ 
is kept fixed in the large-$N$ limit 
and consequently also $\mu$ from eq. (\ref{mutau}) is kept finite and 
unscaled, in contrast to eq. (\ref{mual}).  The eigenvalues are now rescaled 
differently, namely with the square root of $N$ \cite{FKS,A01}, 
\be
\sqrt{N}\,(\re\ z+i\im\ z)\ =\ \sqrt{N}\,z \ \equiv \xi \ \ \ .
\label{microstrong}
\ee
This defines our microscopic origin limit at strong non-Hermiticity. 
In consequence we also have to change the rescaling of the microscopic 
kernel and correlation functions accordingly, 
\bea
K^{(a)}_S(\xi_1,\xi_2^\ast) &\equiv&  \lim_{N\to\infty}\frac{1}{N}\ 
K^{(a)}_N\left(\frac{\xi_1}{\sqrt{N}},\frac{\xi_2^\ast}{\sqrt{N}}\right), \nn\\
\rho_S^{(a)}(\xi_1,\ldots,\xi_k)&\equiv&  \lim_{N\to\infty}\frac{1}{N^{k}}\ 
\rho_N^{(a)}\left(\frac{\xi_1}{\sqrt{N}},\ldots,\frac{\xi_k}{\sqrt{N}}\right).
\label{strongmicrorho+ker}
\eea
In the strong scaling limit eq. (\ref{microstrong}) the argument of the 
Laguerre polynomials in eq. (\ref{kernelN}) can be rewritten in terms of the 
scaling variable $\xi$ alone. In particular we cannot use the asymptotic 
eq. (\ref{asymp}) for the Laguerre polynomials any longer. 
Fortunately there is an identity for an infinite sum of Laguerre polynomials 
at hand \cite{grad}, 
\be
\sum_{k=0}^\infty \frac{k!}{\Gamma(a+1+k)} \tau^{2k}
L_k^a(Z_1^2) L_k^a(Z_2^{\ast2}) \ =\ 
\frac{(Z_1^2Z_2^{\ast2}\tau^2)^{-\frac{a}{2}}}{1-\tau^2}
\exp\left[\frac{-\tau^2}{1-\tau^2}(Z_1^2+Z_2^{\ast2})\right]
I_a\left(2\frac{\sqrt{Z_1^2Z_2^{\ast2}\tau^2}}{1-\tau^2}\right),
\label{LMehler}
\ee
which is the analog of Mehlers formula for the Hermite polynomials used 
in \cite{FKS} to derive the strong correlation functions. 
We can thus immediately read off the large-$N$ limit of the kernel 
eq. (\ref{kernelN}) and obtain
\bea
K_S^{(a)}(\xi_1,\xi_2^\ast)&=&
\frac{2^a\ \Gamma(a+1)}{
\pi \Gamma\left(a+\frac32\right)
(1-\tau^2)^{\frac{a}{2}+\frac74} 
P_{a+\frac32}\left(\frac{1}{\sqrt{1-\tau^2}}\right)
}\ 
\frac{|\xi_1\xi_2^\ast|^{a+\frac12}}{(\xi_1\xi_2^\ast)^a} 
\label{strongkernel}\\
&&
\times\exp\left[
\frac{-1}{2(1-\tau^2)}\left(|\xi_1|^2 + |\xi_2^\ast|^2
-\frac{\tau}{2}({\xi^\ast_1}^2 -\xi_1^2 +\xi_2^2
-{\xi^\ast_2}^2)\right)
\right]
I_a\left(\frac{\xi_1\xi_2^\ast}{1-\tau^2}\right).
\nn
\eea
Here we have inserted the explicit expression for the normalization integral 
eq. (\ref{fdef}).
After taking out factors from the determinant the 
correlation functions again simplify considerably and our final result reads
\be
\rho^{(a)}_S(\xi_1,\ldots,\xi_k) =\prod_{l=1}^k
\left(\frac{2^a\ \Gamma(a+1)\ 
|\xi_l|}{
\pi \Gamma\left(a+\frac32\right)
(1-\tau^2)^{\frac{a}{2}+\frac74} 
P_{a+\frac32}\left(\frac{1}{\sqrt{1-\tau^2}}\right)
}
\ \mbox{e}^{-\frac{1}{1-\tau^2}|\xi_l|^2}
\right)
\!\det_{1\leq i,j\leq k}\left[
I_a\left(\frac{\xi_i\xi_j^\ast}{1-\tau^2}\right)
\right].
\label{strongrho}
\ee
In ref. \cite{Janco} correlation functions for a charged 
two-dimensional plasma with fractional charges at the origin have been 
obtained, in what we call strong non-Hermiticity limit. However, 
since the interaction term induced by the Vandermonde determinant of 
squared arguments in our 
model eq. (\ref{Zev}) is different from that of \cite{Janco} the correlation 
functions obtained there also differ from ours. 

A consistency check for the correlation functions eq. (\ref{strongrho}) 
can be obtained from 
the weakly non-Hermitian correlations eq. (\ref{weakrho}) by taking the limit 
$\al\to\infty$ there. We evaluate the integral inside the determinant as 
follows, 
\bea
\int_0^1 dt\ \mbox{e}^{-\al^2t} 
J_a(\sqrt{2t}\xi_i)J_a(\sqrt{2t}\xi_j^\ast) &=& 
\frac{1}{\al^2}\int_0^\infty ds \mbox{e}^{-s} 
J_a\left(\frac{\sqrt{2s}}{\al}\xi_i\right)
J_a\left(\frac{\sqrt{2t}}{\al}\xi_j^\ast\right) \nn\\
&=& \frac{1}{\al^2} \exp\left[-\frac{1}{2\al^2} (\xi_1^2+\xi_2^{\ast2})\right]
I_a\left( \frac{\xi_1\xi_2^\ast}{\al^2}\right) 
\eea
When identifying $\al^2=1-\tau^2$ and taking out the exponential of the 
determinant in eq. (\ref{weakrho}) we immediately recover the 
strong correlations eq. (\ref{strongrho}). In 
order to match the normalizations we also have to expand the Legendre function 
for large argument, according to eq. (\ref{Legendre}).
We have thus control over the full parameter range starting from real 
eigenvalues via weak to strong non-Hermitian correlation functions.

As an example for eq. (\ref{strongrho}) 
we depict the strong microscopic density in Fig. 
\ref{strongplot} for zero and one massless flavors. The value for $\tau=0.5$ 
corresponds to $\mu=\sqrt{3}/4\approx 0.43$ in our units, according to eq. 
(\ref{mutau}).
The microscopic spectral density develops a hole at the origin and becomes 
flat for large values. This has indeed been observed in quenched lattice data
for intermediate values of the chemical potential \cite{MPW}. 
For one massless flavor ($a=1$) the level repulsion at the origin is again 
stronger and the barrier gets flattened out. The picture is very much 
reminiscent of the microscopic density with massless flavors 
of the non-chiral model \cite{A01} in the strong limit (compared to Fig. 3 
there). 
\begin{figure}[-h]
\centerline{\epsfig{figure=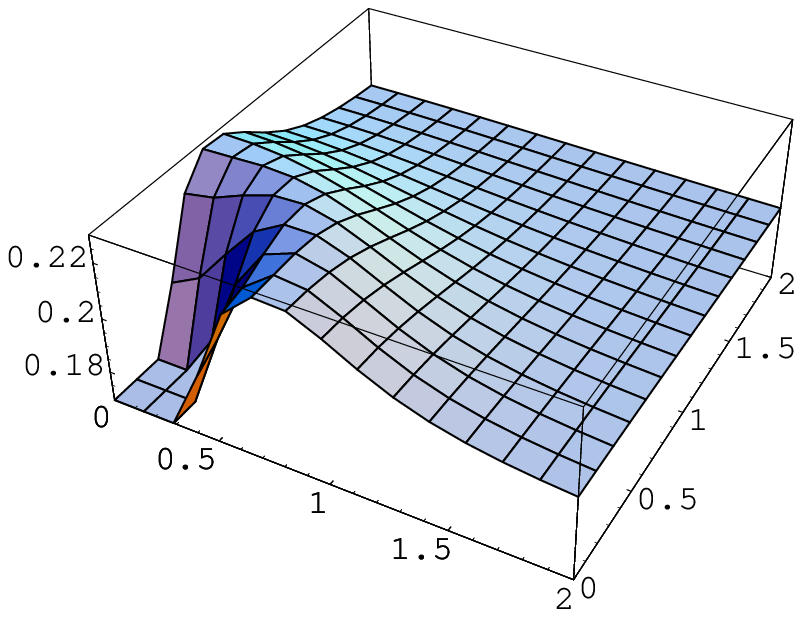,width=20pc}
\put(-160,20){$\re\ \xi$}
\put(-30,50){$\im\ \xi$}
\put(-240,140){$\rho_s^{(0)}(\xi)$}
\put(80,20){$\re\ \xi$}
\put(210,50){$\im\ \xi$}
\put(0,140){$\rho_s^{(1)}(\xi)$}
\epsfig{figure=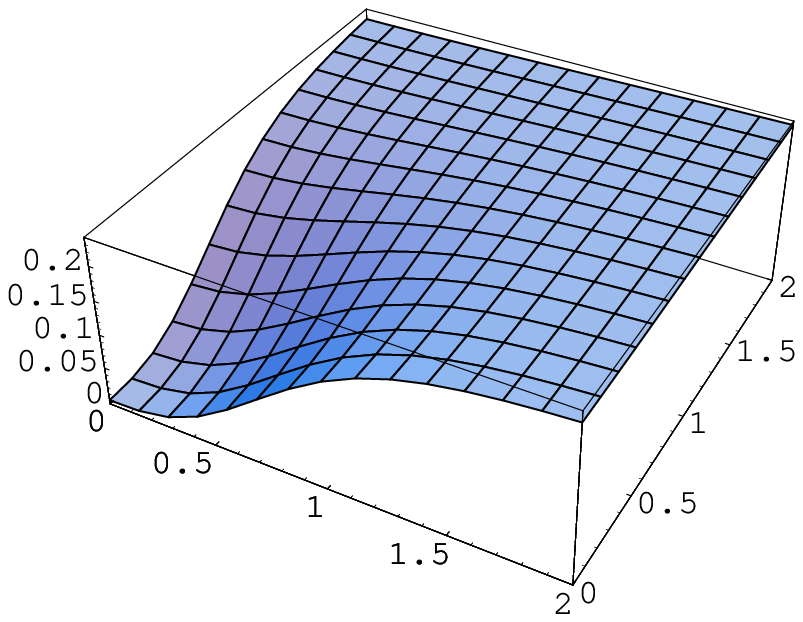,width=20pc}}
\caption{
The microscopic density at $\tau=0.5$ for $a=0$ (left) and $a=1$ (right) 
massless flavors. 
}
\label{strongplot}
\end{figure}
When getting closer to the Hermitian limit by raising $\tau\nearrow1$ 
or equivalently lowering $\mu\searrow0$ 
the repulsion at the origin gets much more pronounced. 
In Fig. \ref{strongplot2} we have depicted the same densities as in Fig. 
\ref{strongplot} but for $\tau=\sqrt{21}/5\approx 0.9$ corresponding to 
$\mu=0.2$. Both values for $\mu$ in Figs. \ref{strongplot} and 
\ref{strongplot2} are in the broken phase when comparing 
to ref. \cite{Steph}. There, the phase transition occurs at a critical value 
$\mu_c=1$ as it follows from eq. (\ref{bound}). 
Even if we take the extreme case of maximal non-Hermiticity which corresponds 
to $\tau=0$ in our model we will stay inside the phase with broken chiral 
symmetry. This can be seen when assuming that the relation $(\ref{mutau})$ 
is still approximately valid, leading to $\mu(\tau=0)=\frac12$. 
\begin{figure}[-h]
\centerline{\epsfig{figure=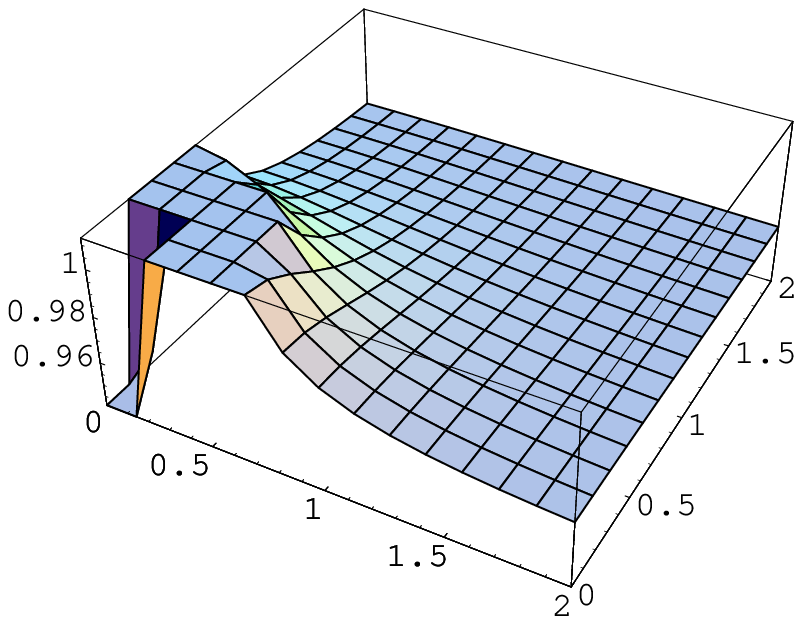,width=20pc}
\put(-160,20){$\re\ \xi$}
\put(-30,50){$\im\ \xi$}
\put(-240,140){$\rho_s^{(0)}(\xi)$}
\put(80,20){$\re\ \xi$}
\put(210,50){$\im\ \xi$}
\put(0,140){$\rho_s^{(1)}(\xi)$}
\epsfig{figure=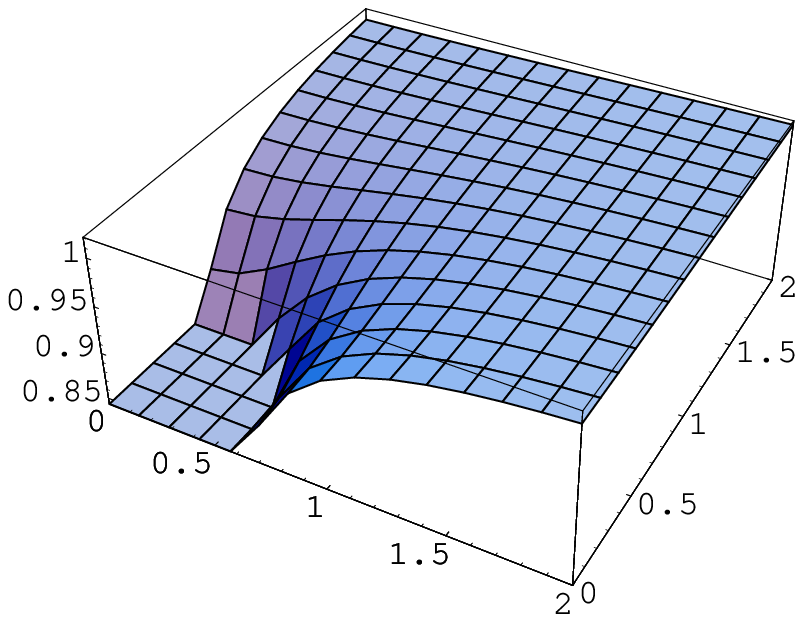,width=20pc}}
\caption{
The microscopic density at $\tau\approx 0.9$ 
for $a=0$ (left) and $a=1$ (right) 
massless flavors. 
}
\label{strongplot2}
\end{figure}

\newpage
\sect{Conclusions}\label{con}   

We have introduced and solved a new class of chiral RMM with complex 
eigenvalues, that corresponds to the extension of the chGUE into the 
complex plane. It was shown that the orthogonal polynomials of our model 
are given by the Laguerre polynomials in the complex plane. Consequently 
all $k$-point correlation functions could be given exactly for a finite number
of eigenvalues $N$ in terms of the kernel of these polynomials. 
We then investigated two different large-$N$ limits with weak and strong 
non-Hermiticity. In both cases explicit results were given for all 
correlation functions in terms of the two different limiting kernels. 
Furthermore, we could prove that the results for weak non-Hermiticity 
interpolate between the known correlation functions of real eigenvalues 
of the chGUE and those of strong non-Hermiticity, taking the limit of the 
weak non-Hermiticity parameter $\al\to 0$ or $\infty$, respectively.

As an application of our model we have proposed that it describes the 
correlations of QCD Dirac eigenvalues close to the origin in the 
presence of a small chemical potential $\mu$. 
For that purpose we related our parameter 
$\tau$ that measures the non-Hermiticity with the chemical potential as it 
occurs in QCD. In the identification we compared to another matrix model 
that incorporates the symmetries of the Dirac operator with $\mu\neq 0$ 
more closely. 
It remains to compare our predictions to lattice data in four-dimensional 
QCD yet to be generated and to see if the success of RMM in predicting 
real eigenvalues extends to the complex plane.
It would be also 
very interesting to find other applications too, in particular 
given the interpretation of the complex eigenvalues as positions of charges 
distributed in a two-dimensional plane. 

So far we have only defined our model as an eigenvalue model and obtained its 
chiral interpretation by analogy with 
the corresponding real eigenvalue model, the chGUE, and its 
non-chiral counterpart, the GUE. 
It would be very interesting to find an explicit matrix 
representation of our model, 
in particular in view of the definition of Dirac mass terms 
and the corresponding solution.

The structure of RMM with complex eigenvalues seems to be much richer as the 
number of symmetry classes is much larger than those of 
real eigenvalue models. While not all of these model may be accessible 
to the technique of orthogonal polynomials presented here at least the chiral
versions of the non-Hermitian or non-symmetric matrices with Dyson index 
$\beta=4$ or $\beta=1$, respectively, could be worth investigating. 
The fact that 
our chiral model with $\beta=2$ again contains the classical polynomials 
of Laguerre type may be of help.

\indent

\noindent
\underline{Acknowledgments}: 
P. Di Francesco and E. Kanzieper are thanked for useful conversation and 
correspondence as well as P. Forrester and M.L. Mehta for bringing several 
references to my attention.
This work was supported by the European network on ``Discrete Random 
Geometries'' HPRN-CT-1999-00161 (EUROGRID).


\end{document}